\documentclass[journal, final]{IEEEtran}
\usepackage{subfigure}
\usepackage[dvips]{graphicx}
\usepackage{amsmath}
\usepackage{cite}
\usepackage{gensymb}
\usepackage{color}
\usepackage{graphicx}
\begin{document}

\title{Differential Bandpass Filters Based on Dumbbell-Shaped Defected Ground Resonators}

\author{Amir~Ebrahimi,~\IEEEmembership{Member,~IEEE,}
        Thomas~Baum,~\IEEEmembership{Member,~IEEE,}
        and Kamran~Ghorbani,~\IEEEmembership{Member,~IEEE}
\thanks{The authors are with the School of Engineering, RMIT University, Melbourne, VIC 3001, Australia (e-mail: amir.ebrahimi@rmit.edu.au).}} %

\maketitle

\begin{abstract}
This letter presents a dumbbell-shaped defected ground resonator and its application in the design of differential filters. The operation principle of the dumbbell-shaped resonator (DSR) coupled to differential microstrip lines is studied through a circuit model analysis. The proposed circuit model is validated through the comparison with the electromagnetic simulation results. It is shown that the bandpass configuration of microstrip-line-coupled DSR can be used to design higher order bandpass filters. The design procedure is explained by developing a third-order filter prototype. The designed filter shows more than $57$~dB common mode rejection within the differential passband. 
\end{abstract}
\IEEEpeerreviewmaketitle
\vspace{-5mm}
\section{Introduction}
\IEEEPARstart{C}{ommon} mode (CM) noise supression is a subject of interest in high-speed RF analog and mixed-mode circuits and systems. The common mode noise degrades the differential mode (DM) signal and power integrity of such systems. One approach for suppressing the common mode noise and electromagnetic interference (EMI) is using a series combination of wideband differential transmission lines and RF filters \cite{Wu2009,Naqui2012}. However, this causes a larger component size that is not desirable in compact systems. More compact sizes can be achieved by designing balanced filters with inherent common mode rejection response \cite{Abbosh2011,Wu2007,Wang2014,Fern'andez-Prieto2015a}. 

Recently, the complementary resonators have shown promising compatibilities in designing compact filters \cite{Ebrahimi2014,Ebrahimi2016}. Variety of differential bandpass filters have been proposed based on the complementary resonators \cite{Velez2013,Horestani2014,Fern'andez-Prieto2015,Velez2016}. In \cite{Velez2013}, the open split-ring resonators (OSRRs) and open complementary split-ring resonators (OCSRRs) are used in designing wideband balanced bandpass filters. S-shaped split-ring resonators have been proposed in \cite{Horestani2014} for designing higher-order differential bandpass filters with compact size. A modified ground plane is considered for enhancing the common mode rejection of a folded step impedance  bandpass filter in \cite{Fern'andez-Prieto2015}. In \cite{Velez2016}, a dual-band bandpass filter is designed by using OCSRRs in combination with compact series \textit{LC} resonators. 

This article presents a new design of balanced bandpass filter based on the dumbbell-shaped resonators (DSRs) embedded in the ground plane of the differential transmission lines. The DSRs provide the possibility of designing higher-order DM filters. Furthermore, in comparison with the similar approaches such as S-shaped CSRRs that have similar circuit models in CM and DM, the DSRs-based filter proposed here offer improved common-mode rejection by showing two distinct equivalent circuits in DM and CM. The next sections describe the operation principle of the DSRs-based differential lines and their application in designing the balanced bandpass filters.

\vspace{-8mm}

\section{Operation Principle}
\label{sec:Basics}
\begin{figure}[!t]
\centering
\includegraphics[width=3.2in]{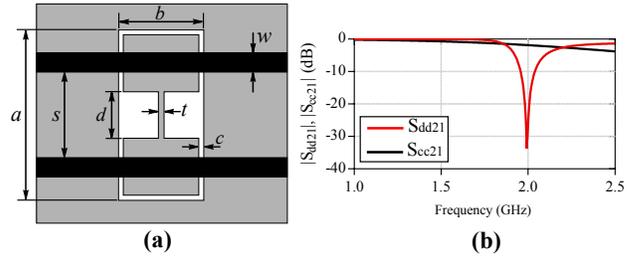}
\caption{(a) Differential transmission lines loaded with DSR. (b) The simulated transmission responses of the structure. The dimensions are: $a=15$~mm, $b=6.8$~mm, $c=t=0.2$~mm, $d=4.2$~mm, $s=5.7$~mm, and $w=0.5$~mm.}
\label{Fig1}
\end{figure}

A typical configuration of a dumbbell-shaped resonator (DSR) loaded differential transmission line is indicated in Fig.~\ref{Fig1}(a). The DSR embedded in the ground plane is composed of two capacitive square patches connected to each other through a thin metallic strip. In this configuration, the resonator will be excited in the differential mode, where the vertical components of the electric fields from the transmission lines are contra-directional. This induces an electric dipole moment across the top and bottom patches of the resonator causing a current flow in the metallic strip between them. However, in common mode, the electric fields of the transmission lines equally excite the top and bottom capacitive patches. Thus, there is no dipole moment and no current flow on the metallic strips and the resonator cannot be excited. This is shown by the EM simulation of the structure in \textit{ADS Momentum} software. 
As seen in Fig.~\ref{Fig1}(b), the resonator causes a bandstop notch in the differential transmission response, whereas the differential line is roughly transparent for the common mode. The substrate in the EM simulation is a $0.635$~mm thick Rogers \textit{RO3010} with $\epsilon_\textrm{r}=10.2$. 

\begin{figure}[!t]
\centering
\includegraphics[width=3.1in]{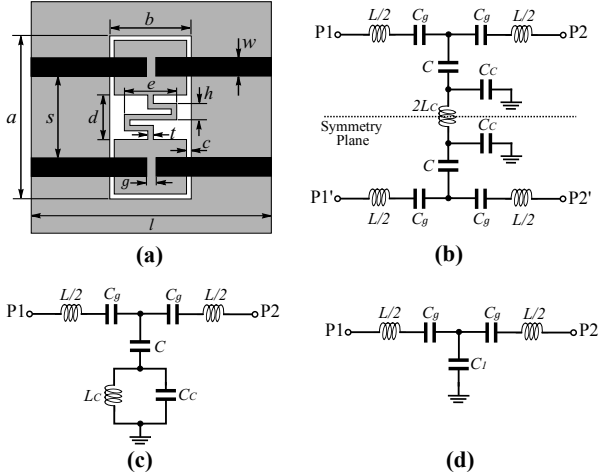}
\caption{(a) Layout of the bandpass DSR-based differential structure. (b) Its equivalent circuit model. (c) Circuit model in differential mode. (d) Circuit model in common mode. The geometrical dimensions are: $a=15.4$~mm, $b=7.6$~mm, $c=0.2$~mm, $d=4.2$~mm, $e=6.4$~mm, $g=0.6$~mm, $h=1.4$~mm, $l=14.4$~mm, $s=5.7$~mm, $t=0.4$~mm, and $w=0.5$~mm.}
\label{Fig2}
\end{figure}

The bandstop response of the DSR-loaded differential line is associated with the negative permittivity of the structure in the vicinity of the DSR resonance. A bandpass behavior will be obtained by adding series capacitive gaps in the microstrip lines as shown in Fig.~\ref{Fig2}(a).
\floatsep 4pt
\begin{figure}[!t]
\centering
\includegraphics[width=2in]{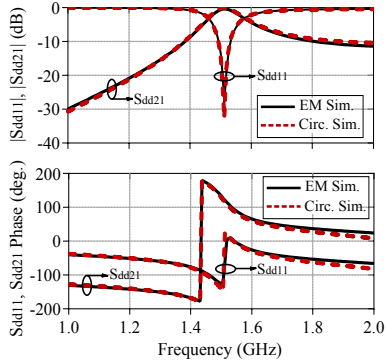}
\caption{Comparison between the S-parameters magnitude and phase obtained from the EM and circuit model simulations of the bandpass structure in differential mode. The circuit element values are: $L=7.4$~nH, $C_g=0.9$~pF, $C=217.5$~pF, $L_C=0.8$~nH, and $C_C=13$~pF.}
\label{Fig3}
\end{figure}
\textfloatsep 4pt
Here, the metallic strip between the patches is meandered for compactness. It should be noted in the design that meandering the thin strip truncates the symmetry resulting in a respectively higher CM to DM (or vice versa) conversion. The circuit model of this structure is presented in Fig.~\ref{Fig2}(b), where the $C_C$ capacitors represent the capacitive effect between the square patches of the DSR and the surrounding ground plane. The 2$L_C$ is the inductance of the metallic strip connecting the square patches to each other, and $C$ is the coupling capacitance between the microstrip lines and the DSR. The $C_g$ capacitors are the series capacitive gaps in the microstrip lines and $L$ is the equivalent inductance of the microstrip lines. In the differential mode, the symmetry plane is a virtual ground and the circuit model takes the form shown in Fig.~\ref{Fig2}(c). In the common mode, the symmetry plane acts as an open circuit and hence, the equivalent circuit for this mode will be the one indicated in Fig.~\ref{Fig2}(d), where $C_\mathrm 1$ is a series combination of $C$ and $C_C$. 

\begin{figure}[!t]
\centering
\includegraphics[width=2in]{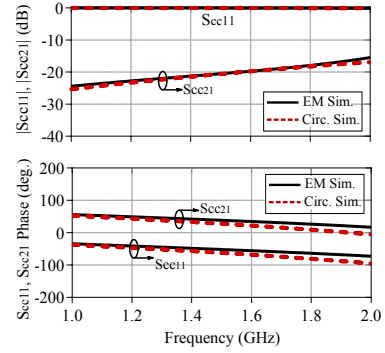}
\caption{The S-parameters magnitude and phase obtained from the EM and circuit model simulations of the bandpass structure in common mode. The circuit element values are: $L=6$~nH, $C_g=1$~pF, and $C_1=12$~pF.}
\label{Fig4}
\end{figure}

The DM equivalent circuit element values are extracted using the procedure explained in \cite{Bonache2006b}. For the CM, initially, the element values are taken from the DM then, more accurate values are obtained by the curve fitting with the EM simulation results in CM. The EM and the circuit model simulation results of the DM and CM are plotted in Fig.~\ref{Fig3} and Fig.~\ref{Fig4} respectively.  The unit cell dimensions are given in the caption of Fig.~\ref{Fig2}, whereas the circuit element values are provided in Fig.~\ref{Fig3} and Fig.~\ref{Fig4}. The good agreement between the EM and circuit simulations verifies the developed circuit model. The differential mode shows a passband around $1.5$~GHz, whereas the common mode transmission is smaller than $-20$~dB around this frequency. It is expected that by a series combination of the unit cells in a filter design, the common mode rejection increases by multiple orders of $20$~dB.

\vspace{-3mm}
\section{Differential BPF Design}
\label{FilterDes}
For validating the potential of the DSR-loaded differential lines in the design of differential bandpass filters, a third-order bandpass filter is designed with a periodic arrangement (periodicity is not mandatory in general) of bandpass DSR-loaded differential lines. The filter is designed with a center frequency of $1.5$~GHz and $6$\% fractional bandwidth. The focus is on the design of the differential mode filtering characteristic since the structure inherently suppresses the common mode.  

The basic bandpass cell in Fig.~\ref{Fig2} is adopted to be used as the filter unit cell. To this end, the element values of the bandpass cell in Fig.~\ref{Fig2}(c) should be designed first. The fractional bandwidth of the periodic filter is obtained as

\begin{equation}
\mathrm{FBW}=\dfrac{\omega_2-\omega_1}{\omega_0}=\dfrac{g\Delta}{2},
\end{equation}
where $g$ is the element values of the lowpass filter prototype, $\Delta$ is the $3$-dB bandwidth of the DSRs, $\omega_0$, $\omega_1$, and $\omega_2$ are the center and the $3$-dB angular frequencies of the filter respectively. The $g$ value for a third-order periodic filter is $1.521$ that corresponds to $\Delta=7.9$\% in the DSRs for $6$\% FBW \cite{Bonache2006a}. By setting $\omega_1$ and $\omega_2$ to be equidistant from $\omega_0$ and assuming a narrowband reponse, the shunt impedance made of $C$, $L_C$, and $C_C$ becomes $jZ_0$, $jZ_0/2$, and infinity at $\omega_0$, $\omega_1$, and $\omega_2$ frequencies respectively, where $Z_0=50~\mathrm{\Omega}$ is the reference impedance \cite{Bonache2006a}. This means

\begin{equation}
\dfrac{1-\omega_0^2 L_C (C+C_C)}{jC\omega_0(1-\omega_0^2 L_C C_C)}=jZ_0,
\end{equation}

\begin{equation}
\dfrac{1-\omega_1^2 L_C (C+C_C)}{jC\omega_1(1-\omega_1^2 L_C C_C)}=jZ_0/2,
\end{equation}

\begin{equation}
1-\omega_2^2 L_C C_C=0.
\end{equation}

The $C$, $C_C$, and $L_C$ values are determined by solving the above equations, while the $C_g$ value is given by

\begin{equation}
C_g=\dfrac{1}{2Z_0\omega_0}.
\end{equation}

Finally, $L$ is obtained by knowing that at $\omega_0$, in $\left| \mathrm{S_{21}}\right|=1$, and $\phi=90^\circ$, in each filter cell \cite{Bonache2006a}. This results in  

\begin{equation}
L=\dfrac{2(1+\omega_0C_gZ_0)}{\omega_0^2 C_g}.
\end{equation}

\floatsep 4pt

\begin{figure}[!t]
\centering
\includegraphics[width=3.2in]{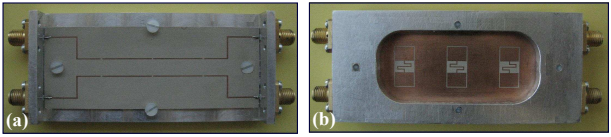}
\caption{The fabricated filter prototype. (a) Front view. (b) Back view.}
\label{Fig5}
\end{figure}

\textfloatsep 4pt

\begin{figure}[!t]
\centering
\includegraphics[width=3in]{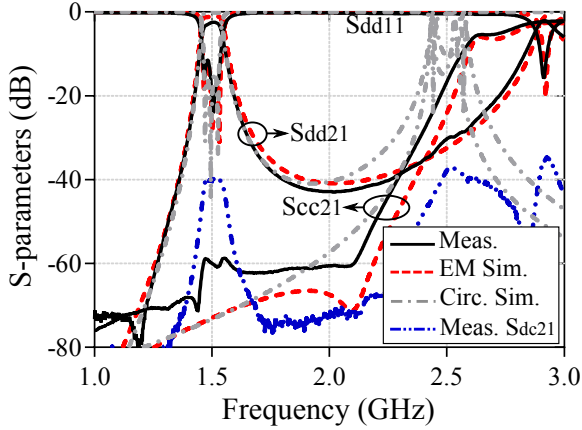}
\caption{Comparison between the measured and simulation results of the filter. }
\label{Fig6}
\end{figure}

Now, the layout dimensions should be optimized for implementation of the designed elements. This is performed by the curve fitting between the EM and the circuit model simulations, where the initial dimensions for inductive traces and capacitive gaps are obtained using the formulas in \cite{Hong2004}.  The optimized dimensions are given in Fig.~\ref{Fig2} except the feed length that is $l=19.8$~mm.  

The photograph of the fabricated filter is shown in Fig.~\ref{Fig5}. An aluminum frame is designed protecting the filter from bending and possible damage during the measurement. The filter size is $0.6\lambda_g \times 0.2\lambda_g$, where $\lambda_g$ is the guided wavelength at $1.5$~GHz. The filter size can be reduced by symmetrically meandering the microstrip lines between the unit cells. Mitered feeds should be used for the filter design at higher frequencies. Fig.~\ref{Fig6} shows the simulated and measured DM and CM responses, where the metallic frame is considered in EM simulations. The common mode rejection is more than $57$~dB within the DM passband. A small passband in CM is a result of asymmetry caused by the fabrication tolerance. The passband insertion loss is $2.4$~dB that can be improved by choosing a thicker substrate resulting in wider microstrip traces with a lower Ohmic loss. A comparison between the proposed filter and the state-of-art structures in Table~\ref{table1} shows a competitive performance of the proposed filter over the previous works.   

\begin{table}
  \renewcommand{\arraystretch}{1.2}
  \caption{Comparison of Various Differential Bandpass Filters} 
  \label{table1}
  \centering
  \begin{tabular}{c@{\hskip 0.05cm} |c@{\hskip 0.05cm}| c@{\hskip 0.1cm} |c@{\hskip 0.03cm} |c@{\hskip 0.1cm} |c@{\hskip 0.02cm}| c@{\hskip 0.05cm}  }
  \hline
   Ref. & \begin{tabular}[x]{@{}c@{}}Order \\ (N) \end{tabular} & \begin{tabular}[x]{@{}c@{}}CMRR in \\Diff. Passband\end{tabular}   & FBW & \begin{tabular}[x]{@{}c@{}}$\left| \mathrm{S_{cc21}} \right|\geq$\\ 30~dB\end{tabular}  & IL~(dB) & \begin{tabular}[x]{@{}c@{}}Size \\ ($\lambda_g^2$) \end{tabular}\\
  \hline
  \cite{Wu2007} & 4 & 31~dB & 12\% & 0.5$f_{0\mathrm{d}}$--5.8$f_{0\mathrm{d}}$& $3.5$& $0.047$\\
  \cite{Wang2014} & 2 & 41~dB& 6.6\% & $f_{0\mathrm{d}}$--1.28$f_{0\mathrm{d}}$& $0.8$& $0.29$\\
  \cite{Fern'andez-Prieto2015a}& 2 & 38.8~dB & 11.1\%& 0--1.67$f_{0\mathrm{d}}$ & $1.28$& $0.048$ \\
\cite{Velez2013} & 3 & 20--50~dB & 45\% & 0.5$f_{0\mathrm{d}}$--1.1$f_{0\mathrm{d}}$& $1.76$& $0.045$ \\
\cite{Horestani2014} & 3 & 25~dB & 10\% & 0.85$f_{0\mathrm{d}}$--1.1$f_{0\mathrm{d}}$& $2.8$& $0.025$ \\
\cite{Fern'andez-Prieto2015} & 2 & 34~dB & 10\% & 0.8$f_{0\mathrm{d}}$--1.2$f_{0\mathrm{d}}$& $1.2$& $0.031$ \\
  T. W. & 3 & 57~dB& 6\% & 0--1.55$f_{0\mathrm{d}}$& $2.4$& $0.12$\\
  \hline 
  \end{tabular}
  \end{table}




\vspace{-4mm}
\section{Conclusion}
Dumble-shaped resonators loaded with differential transmission lines have been proposed for the design of balanced banpass filters. The electromagnetic response of the unit cell has been analyzed using the equivalent circuit model. A designed and fabricated third-order balanced bandpass filter based on DSRs offers an in-band common-mode rejection ratio of $57$~dB.


\end{document}